\documentclass[%
 reprint,
 amsmath,amssymb,
 aps,twocolumn
]{revtex4}

\usepackage{natbib}
\usepackage{aas_macros}
\usepackage{graphicx}
\usepackage{dcolumn}
\usepackage{bm}

\newcommand\ba{\begin{eqnarray}}
\newcommand\ea{\end{eqnarray}}

\usepackage{color}

 \newcommand\vbc{{\bf v}_{bc}}
 
\begin{document}


\title{Impact of dark matter-baryon relative velocity on the 21cm forest}

\author{Hayato Shimabukuro}
 \affiliation{Yunnan University, SWIFAR, No.2 North Green Lake Road, Kunming, Yunnan Province,650500, China\\
 Graduate School of Science, Division of Particle and Astrophysical Science, Nagoya University, Chikusa-Ku, Nagoya, 464-8602, Japan}
  
 \email{shimabukuro@ynu.edu.cn}
\author{Kiyotomo Ichiki}
 \affiliation{%
Graduate School of Science, Division of Particle and Astrophysical Science, Nagoya University, Chikusa-Ku, Nagoya, 464-8602, Japan
}%
 \affiliation{%
 Kobayashi-Maskawa Institute for the Origin of Particles and the Universe, Nagoya University, Chikusa-ku, Nagoya, 464-8602, Japan
 }%

 \affiliation{%
 Institute for Advanced Research, Nagoya University,  Furocho,
Chikusa-ku, Nagoya, Aichi 464-8602, Japan
 }%
  \email{ichiki.kiyotomo@c.mbox.nagoya-u.ac.jp}
\author{Kenji Kadota}
\affiliation{
  School of Fundamental Physics and Mathematical Sciences,
  Hangzhou Institute for Advanced Study,\\
  University of Chinese Academy of Sciences (HIAS-UCAS), Hangzhou 310024, China\\
  International Centre for Theoretical Physics Asia-Pacific (ICTP-AP), Beijing/Hangzhou, China
}
 \email{kadota@ucas.ac.cn}

\date{\today}

\begin{abstract}
We study the effect of the relative velocity between the dark matter (DM) and the baryon on the 21cm forest signals.
The DM-baryon relative velocity arises due to their different evolutions before the baryon-photon decoupling epoch and it gives an additional anisotropic pressure that can suppress the perturbation growth. 
It is intriguing that the scale $k\sim {\cal O}(10\sim 10^3)h/\mathrm{Mpc}$ at which the matter power spectrum is affected by such a streaming velocity turns out to be the scale at which the 21cm forest signal is sensitive to. We demonstrate that the 21cm absorption line abundance can decrease by more than a factor of a few due to the small-scale matter power spectrum suppression caused by the DM-baryon relative velocity.

\end{abstract}

\maketitle


\section{Introduction}
Over the past decades, the hierarchical structure formation scenario based on the standard cold dark matter($\Lambda$CDM) model has been established by observations such as the cosmic microwave background (CMB) and galaxies that trace the large-scale structure of the universe \citep[e.g. Refs.][]{2020A&A...641A...6P,2021PhRvD.103h3533A}. In particular, the matter power spectrum measured by the cosmological observations is consistent with the $\Lambda$CDM model at large scales. On the other hand, the current Lyman alpha forest observations investigate the fluctuations of the hydrogen atoms in the intergalactic medium (IGM) and constrain the matter power spectrum at small scales up to $k\sim 2 h\mathrm{Mpc}^{-1}$\citep[e.g. Refs.][]{2013A&A...559A..85P,2019MNRAS.489.2247C}. Scales smaller than $\sim 1$ Mpc which the Lyman alpha forest cannot probe have not been well explored observationally, and the 21cm absorption line system (called the 21cm forest, in analogy to the Lyman forest) is a promising tool to explore those smaller scales. The 21cm absorption line is caused by neutral hydrogen atoms due to the hyperfine structure. They appear as absorption lines in the spectra of high redshift radio-loud sources due to neutral hydrogen atoms in the intervening cold neutral IGM and collapsed objects \citep[e.g. Refs.][]{2002ApJ...577...22C,2002ApJ...579....1F,2006MNRAS.370.1867F,2009ApJ...704.1396X,2012MNRAS.425.2988M,2015aska.confE...6C,2016MNRAS.455..962S,2021MNRAS.506.5818S}. 

In particular, the smallest collapsed objects, called ``minihalos'' can the promising sources to lead to the 21cm forest for scales corresponding to $k\gtrsim 10$ [$\mathrm{Mpc^{-1}}$]. The minihalos can form when the virial temperature is below the threshold where the atomic cooling becomes ineffective ($T_{\mathrm{vir}} \lesssim 10^4 \mathrm{K}$). Thus, they cannot cool effectively and cannot collapse to form protogalaxies. In such a condition, the masses of the minihalos are of order $M\lesssim 10^8 M_{\odot}$ corresponding to the scales $k\gtrsim 10$ [$\mathrm{Mpc^{-1}}$]. 

Those small scales are of great interest for further exploration of the $\Lambda$CDM cosmology and its extensions. For instance, there are some challenges at small scales for $\Lambda$CDM cosmology such as the ``missing satellite problem'', the ``core-cusp problem'' and the ``too big to fail problem''(see e.g. Refs. \citep[]{2017ARA&A..55..343B}). The small-scale structures probed by the 21cm forest can also give a clue to the nature of dark matter, such as warm dark matter\citep{2014PhRvD..90h3003S}, ultralight axionlike particles\citep{2020PhRvD.101d3516S,2020PhRvD.102b3522S,2022JCAP...08..066K}, and primordial black holes\citep{2022PASJ..tmp...28V}.

In this paper, we aim to demonstrate that the effects of the relative velocity between the dark matter and the baryon can also be imprinted in the 21cm forest signals.
Such a relative velocity arises due to the different evolution between dark matter and baryons before the photon-baryon decoupling epoch \cite{2010PhRvD..82h3520T}.
The photons and baryons are tightly coupled through Thomson scattering until the recombination epoch when the number density of free electrons becomes low, while the CDM fluctuations keep growing under DM's own gravity. Such a different evolution results in supersonic relative velocity between the baryon and DM at the recombination epoch with the root-mean-square (RMS) velocity $\sigma_{\mathrm{bc}} \sim 30\mathrm{km/s}$ which is coherent over the scales of order a few comoving Mpc, while the sound speed of the baryon fluid drops dramatically to about 6 km/s (and the Jeans length also drops dramatically to $\sim 30$ kpc).
Such a supersonic stream of baryons through dark matter without falling into the potential well of DM can consequently result in the structure formation suppression \cite{2010PhRvD..82h3520T,Yoo:2011tq,2010JCAP...11..007D,Beutler:2016zat,Fialkov:2012su,Visbal:2012aw}.

The relevant scales sensitive to relative velocity are expected to be around the baryon Jeans scales at which the fluctuations are sensitive to both baryonic pressure and gravitational infall. The larger scale growth is more affected by the gravitational attraction to the potential well, and the smaller scale growth is more affected by the baryonic pressure which exists even without supersonic relative velocity. Hence, the large suppression of the matter power spectrum due to the streaming velocity is expected to appear around the Jeans scale $k$ of order  $ {\cal O}(10^2)\mathrm{Mpc}^{-1}$ at the redshift $z\gtrsim 10$, and it is intriguing that the 21cm forest observations can provide unique and promising probes on such small scales.
This paper focuses on the effects of halo abundance suppression and leaves more detailed numerical calculations, including the change of gas profiles in the presence of the streaming velocity, to the future work \cite{Stacy:2010gg,Greif:2011iv,Naoz:2011if,Richardson:2013uqa,Fialkov:2011iw}.

In this paper, we estimate how the relative velocity impacts the 21cm forest signals. Section \ref{Sec2} gives a brief review of the effects of the DM-baryon relative velocity on the matter power spectrum and the halo mass function. We then outline the formalism to estimate the 21cm absorption line abundance in Section \ref{Sec3}. Section \ref{Sec4} gives our results, followed by a discussion and conclusions in Section \ref{Sec5}.

\section{Impact of relative velocity on structure formation}
\label{Sec2}
\subsection{Relative velocity}

After the recombination, both baryon and CDM perturbations grow as pressureless fluids above the Jeans scale with a characteristic relative velocity $v_{bc}\sim 30 (1+z)/1000$ [km/s] and baryon sound speed $c_s =\sqrt{\gamma k_{\mathrm{B}}T_b/\mu m_{\mathrm{H}}}$. Here, $\gamma=5/3$ for an ideal monoatomic gas; $\mu=1.22$ is the mean molecular weight, including a helium mass fraction of 0.24; and $m_{\mathrm{H}}$ is the mass of hydrogen. 
The density fluctuation evolutions are governed by the following equations  
\ba
\frac{\partial \delta_c}{\partial t}+a^{-1} {\bf v}_c \cdot \nabla \delta_c&=&-a^{-1}(1+\delta_c) \nabla \cdot {\bf v}_c
\label{eq:1}
\\
\frac{\partial \delta_b}{\partial t}+a^{-1} {\bf v}_b \cdot \nabla \delta_b&=&-a^{-1}(1+\delta_b) \nabla \cdot {\bf v}_b
\\
\frac{\partial {\bf v}_c}{\partial t}+a^{-1} \left( {\bf v}_c \cdot \nabla \right) {\bf v}_c&=& -\frac{\nabla \Phi}{a}-H {\bf v}_c
\\
\frac{\partial {\bf v}_b}{\partial t}+a^{-1} \left( {\bf v}_b \cdot \nabla \right) {\bf v}_b&=& -\frac{\nabla \Phi}{a}-H {\bf v}_b-a^{-1}c_s^2 \nabla \delta_b
\\
a^{-2}\nabla^2 \Phi &=& 4\pi G \bar{\rho}_m \delta_m
\label{eq:5}
\ea
which represent the continuity equations for the CDM and baryon density fluctuation ($\delta_c,\delta_b$), the Navier-Stokes equations for the CDM and baryon velocity (${\bf v}_c,{\bf v}_b$), and the Poisson equation for the gravitational potential $\Phi$.
The different evolutions of baryon and CDM fluctuations before the decoupling can source the relative velocity between them, $\vbc$, which can be supersonic just after the decoupling epoch, and the effects on the cosmological observables such as the baryon acoustic oscillations and 21cm fluctuations are discussed \cite{Yoo:2011tq,2010JCAP...11..007D,Beutler:2016zat,Fialkov:2012su,Visbal:2012aw}. In this paper, we discuss how the 21cm forest signals can be modified in the presence of such a streaming velocity, with a particular emphasis on the suppression of the minihalo formation, compared to those estimated without taking account of the non-negligible $\vbc$.
Note that this is a nonlinear effect arising from the quadratic terms in the perturbation evolution equations, and it does not show up in the conventional linear theory \cite{2010PhRvD..82h3520T}. The resultant advection of baryons out of the dark matter potential well caused by the large-scale velocity flow (associated with the baryon-photon fluid acoustic oscillations up to the sound horizon scale $\sim 150$ comoving Mpc) leads to the modulated small-scale perturbation suppression.

\subsection{Matter power spectrum and mass function}
\begin{figure}
    \includegraphics[width=1.0\hsize]{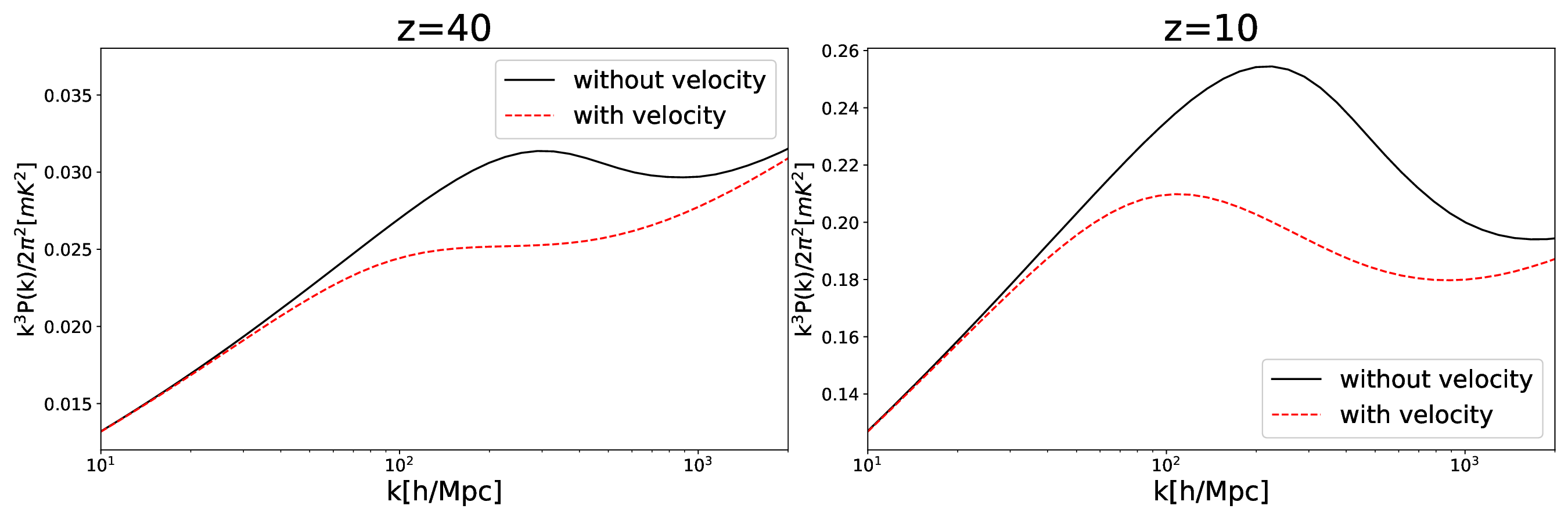}
    \caption{Matter power spectrum at $z=40, 10$ without(solid line) and with(dashed line) the effect of relative velocity.
    }
    \label{fig:ps}
\end{figure}

The presence of relative velocity changes the structure formation of the Universe. To understand the impact of relative velocity, we first note the impact of relative velocity on the matter power spectrum by solving Eqs. (\ref{eq:1})--(\ref{eq:5}). In Fig.\ref{fig:ps}, we show the matter power spectrum, including relative velocity 
at $z=40,10$. At $z=40$, the matter power spectrum is suppressed by $\sim10\%$ at 50 - 500 $\mathrm{Mpc^{-1}}$ and maximally suppressed at the Jeans scale $k_{\mathrm{J}}\sim 200 \mathrm{Mpc}^{-1}$. The Jeans scale is the scale of balance between gravitational attraction and baryonic perturbation; thus additional pressure due to relative velocity impacts the growth of structure formation at these scales. On the other hand, perturbations on much larger scales are not affected by pressure and gravitationally collapse, whereas the pressure at smaller scales prevents perturbations from collapsing. This qualitative argument also applies to the matter power spectrum suppression at $z=10$. 

 \begin{figure}
    \includegraphics[width=0.8\hsize]{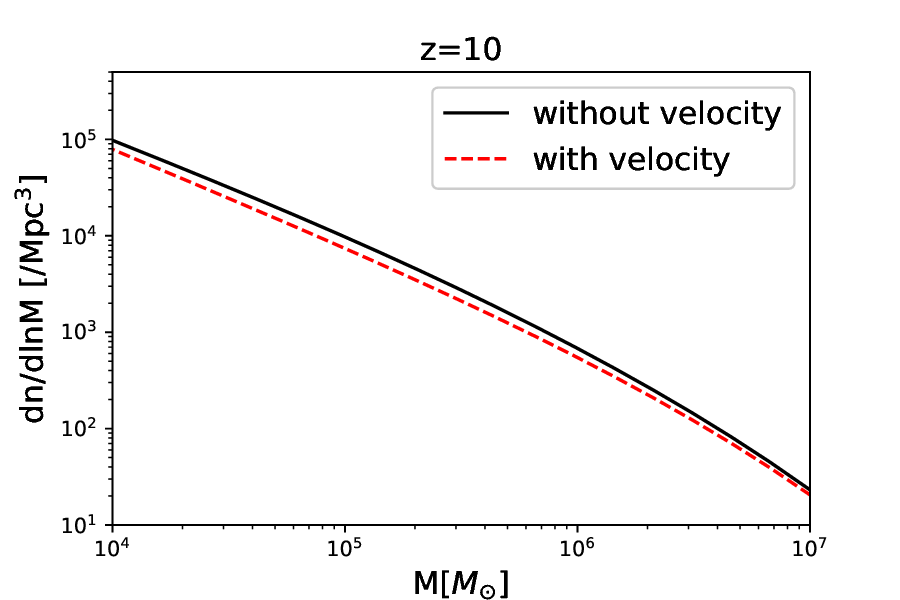}
    \caption{Halo mass abundance at $z=10$ without (solid line) and with (dashed line) the effect of relative velocity.
    }
    \label{fig:MF}
\end{figure}

We next see the impact of relative velocity on the mass function. For the halo mass function, we adopt the Press-Schechter mass function\citep{1974ApJ...187..425P}. We mention that the Sheth-Tormen mass function is more precise at least for a low redshift \citep{1999MNRAS.308..119S}. The halo mass function at a high redshift of our interest is, however,  less certain, and the difference between the Press-Schechter formalism and the Sheth-Tormen formalism is small and does not affect our discussions. Hence, the Press-Schechter mass function suffices for our purpose of demonstrating the potential effects of the relative velocity on the 21cm forest signals. 
Fig.\ref{fig:ps} shows the anticipated suppression in the matter power spectrum in the presence of relative velocity at scales of $k$=20-3000 $\mathrm{Mpc}^{-1}$ at $z$=10. The corresponding halo mass scale enclosed in the sphere $r= 2\pi /k$ is around $ 4.7 \times 10^3 M_{\odot}\lesssim M\lesssim 5.3 \times 10^9 M_{\odot}$\citep[e.g. Ref.][]{2017ARA&A..55..343B}, and the suppression of the abundance of those minihalos is illustrated in Fig. \ref{fig:MF}.

\section{21cm forest}
\label{Sec3}
The 21cm forest is a system of 21cm absorption lines that appear in the continuum spectrum of a radio background source. The 21cm absorption lines are generated by neutral hydrogen atoms in the diffuse IGM or minihalos. The column density of minihalos is larger than that of diffuse IGM and hence minihalos can produce a larger optical depth. In this work, we focus on the 21cm forest generated by neutral hydrogen gas in the minihalos based on \citep{2002ApJ...579....1F,2014PhRvD..90h3003S,2020PhRvD.101d3516S,2020PhRvD.102b3522S}. 

\subsection{Halo and gas profiles}

We first specify the dark matter halo and gas profiles analytically for the estimation of the 21cm absorption lines. We assume the Navarro-Frenk-White (NFW) profile for the dark matter density distribution\citep{1997ApJ...490..493N,2000ApJ...540...39A} characterized by the concentration parameter $y=r_{\rm vir}/r_{s}$,where $r_{s}$ is the scale radius and $r_{\rm vir}$ is the virial radius given by \citep{2001PhR...349..125B}

\begin{equation}
\begin{aligned}
r_{\mathrm{vir}}=& 0.784\left(\frac{M}{10^8 h^{-1} M_{\odot}}\right)^{1 / 3}\left[\frac{\Omega_m}{\Omega_m^z} \frac{\Delta_c}{18 \pi^2}\right]^{-1 / 3} \\
& \times\left(\frac{1+z}{10}\right)^{-1} h^{-1}[\mathrm{kpc}].
\label{eq:virial}
\end{aligned}
\end{equation}
Here, $\Delta_c=18 \pi^2+82 d-39 d^2$ is the overdensity of halos collapsing at redshift $z$ with $d=\Omega_m^z-1$ and $\Omega_m^z=\Omega_m(1+z)^3 /\left(\Omega_m(1+z)^3+\Omega_{\Lambda}\right)$. The concentration parameter $y$ depends on the halo mass and redshift. \citet{2001MNRAS.321..559B} conducted high-resolution N-body simulations and showed that the redshift dependence of the concentration parameter is inversely proportional to $(1+z)$. We assume that the concentration parameter $y$ is given by the fitting formula obtained by \citep{2005MNRAS.363..379G}. Given the dark matter density profile, we can obtain the gas density profile analytically with the assumption that gas is in isothermal and hydrostatic equilibrium states. In that case, the gas density profile is given by \citep{1998ApJ...497..555M, 2011MNRAS.410.2025X}

\begin{equation}
\ln \rho_g(r)=\ln \rho_{g 0}-\frac{\mu m_p}{2 k_{\mathrm{B}} T_{\mathrm{vir}}}\left[v_{\mathrm{esc}}^2(0)-v_{\mathrm{esc}}^2(r)\right],
\label{eq:gas_profile}
\end{equation}
where $\mu=1.22$ is the mean molecular weight of the gas and $m_p$ is the proton mass. Note that $T_{\mathrm{vir}}$ is the virial temperature given by 

\begin{equation}
\begin{aligned}
T_{\mathrm{vir}}=& 1.98 \times 10^4\left(\frac{\mu}{0.6}\right)\left(\frac{M}{10^8 h^{-1} M_{\odot}}\right)^{2 / 3}\left[\frac{\Omega_m}{\Omega_m^z} \frac{\Delta_c}{18 \pi^2}\right]^{1 / 3} \\
& \times\left(\frac{1+z}{10}\right)[\mathrm{K}].
\end{aligned}
\end{equation}
and $\rho_{g0}$ is the central gas density given by 

\begin{equation}
\rho_{g 0}(z)=\frac{\left(\Delta_c / 3\right) y^3 e^A}{\int_0^y(1+t)^{A / t} t^2 d t}\left(\frac{\Omega_b}{\Omega_m}\right) \bar{\rho}_m(z),
\end{equation}
where $A=3 y / F(y)$ and $F(y)=\ln (1+y)-y /(1+y)$.$\bar{\rho}_m(z)$ is the mean total matter density at $z$. Note that $v_{\mathrm{esc}}$ is the escape velocity given by 

\begin{equation}
v_{\mathrm{esc}}^2(r)=2 \int_r^{\infty} \frac{G M\left(r^{\prime}\right)}{r^{\prime 2}} d r^{\prime}=2 V_c^2 \frac{F(y x)+y x /(1+y x)}{x F(y)},
\end{equation}
where $x\equiv r/r_{\mathrm{vir}}$, and $V_{\mathrm{c}}$ is the circular velocity given by 

\begin{equation}
\begin{aligned}
V_c^2=\frac{G M}{r_{\mathrm{vir}}}=& 23.4\left(\frac{M}{10^8 h^{-1} M_{\odot}}\right)^{1 / 3}\left[\frac{\Omega_m}{\Omega_m^z} \frac{\Delta_c}{18 \pi^2}\right]^{1 / 6} \\
& \times\left(\frac{1+z}{10}\right)^{1 / 2}[\mathrm{~km} / \mathrm{s}].
\end{aligned}
\end{equation}

We leave more detailed numerical calculations including the change of the temperature-gas profiles and the star formation rate in the presence of streaming velocity, to future work \cite{Stacy:2010gg,Greif:2011iv,Naoz:2011if,Richardson:2013uqa,Fialkov:2011iw}.

\subsection{Spin temperature}

The spin temperature is a key quantity in estimating the 21cm line spectrum. The spin temperature $T_{\mathrm{S}}$ is defined by the ratio between the number density $n_i$ of the neutral hydrogen atom in the two hyperfine levels (singlet $n_0$ and triplet $n_1$)

\begin{equation}
\frac{n_1}{n_0}=3\exp \left(\frac{-h \nu_{21}}{k_{\mathrm{B}} T_{\mathrm{S}}}\right),
\end{equation}
where $h$ is the Planck constant, $k_{\mathrm{B}}$ is the Boltzmann constant, and $\nu_{21}=1.4$GHz. The spin temperature is determined by the interaction of neutral hydrogen with CMB and Lyman-$\alpha$ photons and by collisions between hydrogen atoms using the following equation (see e.g. Ref. \citep[][]{2023PASJ...75S...1S}),

\begin{equation}
T_{\mathrm{S}}^{-1}=\frac{T_\gamma^{-1}+x_c T_{\mathrm{K}}^{-1}+x_\alpha T_{\mathrm{C}}^{-1}}{1+x_c+x_\alpha}.
\end{equation}

Here, $T_{\gamma}=2.73 \mathrm{K}$ is the CMB temperature at redshift $z$, $T_{\mathrm{K}}$ is the gas kinetic temperature, and $T_{\mathrm{C}}$ is the color temperature of Lyman-$\alpha$ photon. Note that $x_{\alpha}$ and $x_{c}$ are coupling coefficients for the interaction with Lyman-$\alpha$ photon collision with the neutral hydrogen atom, respectively. To understand how the relative velocity impacts the 21cm forest, we ignore any UV radiation field and radiative feedback, and thus we set $x_{\alpha}=0$. We also set $T_{K}=T_{\mathrm{vir}}$. This is a good approximation for a minihalo because the minihalos are considered collapsed objects with $T_{\mathrm{vir}}=T_{\mathrm{K}} < 10^4 \mathrm{K}$ and the gas cooling is inefficient within the minihalos. Note that even if we turn on Wouthuysen-Field(WF) effect ($x_a \neq 0$), its effect on the 21cm forest is small. In the case of $x_a \neq 0$, the spin temperature couples to the gas temperature due to the WF effect, but the 21cm forest is mainly produced by the contribution of the neutral hydrogen atoms inside the minihalo, where the neutral hydrogen density is large; in this region, the spin temperature already collisionally couples to the gas temperature (here, the virial temperature). In the case of $x_a\neq 0$, the spin temperature couples to the gas temperature mainly in the outer regions of the minihalo due to the WF effect, but the contribution to the optical depth from the outer regions of the minihalo to the 21cm forest is small (see, e.g., Refs.\citep[][]{2011MNRAS.410.2025X,2014PhRvD..90h3003S}). Although the main contributions to collisional coupling are H-H and H-$e^{-}$, we only consider H-H collision for computation of $x_{c}$ because the fraction of free electron is small in a minihalo and we can ignore its effect\citep{2005ApJ...622.1356Z,2006MNRAS.370.1867F,2014PhRvD..90h3003S}. The spin temperature approaches the virial temperature in the inner regions of a minihalo(and thus a larger minihalo has a larger spin temperature) and approaches the CMB temperature in the outer regions of a minihalo because the collisional coupling becomes ineffective due to the small gas density \cite{2014PhRvD..90h3003S}.

\subsection{Optical depth}

\begin{figure}
    \centering
    \includegraphics[width=0.8\hsize]{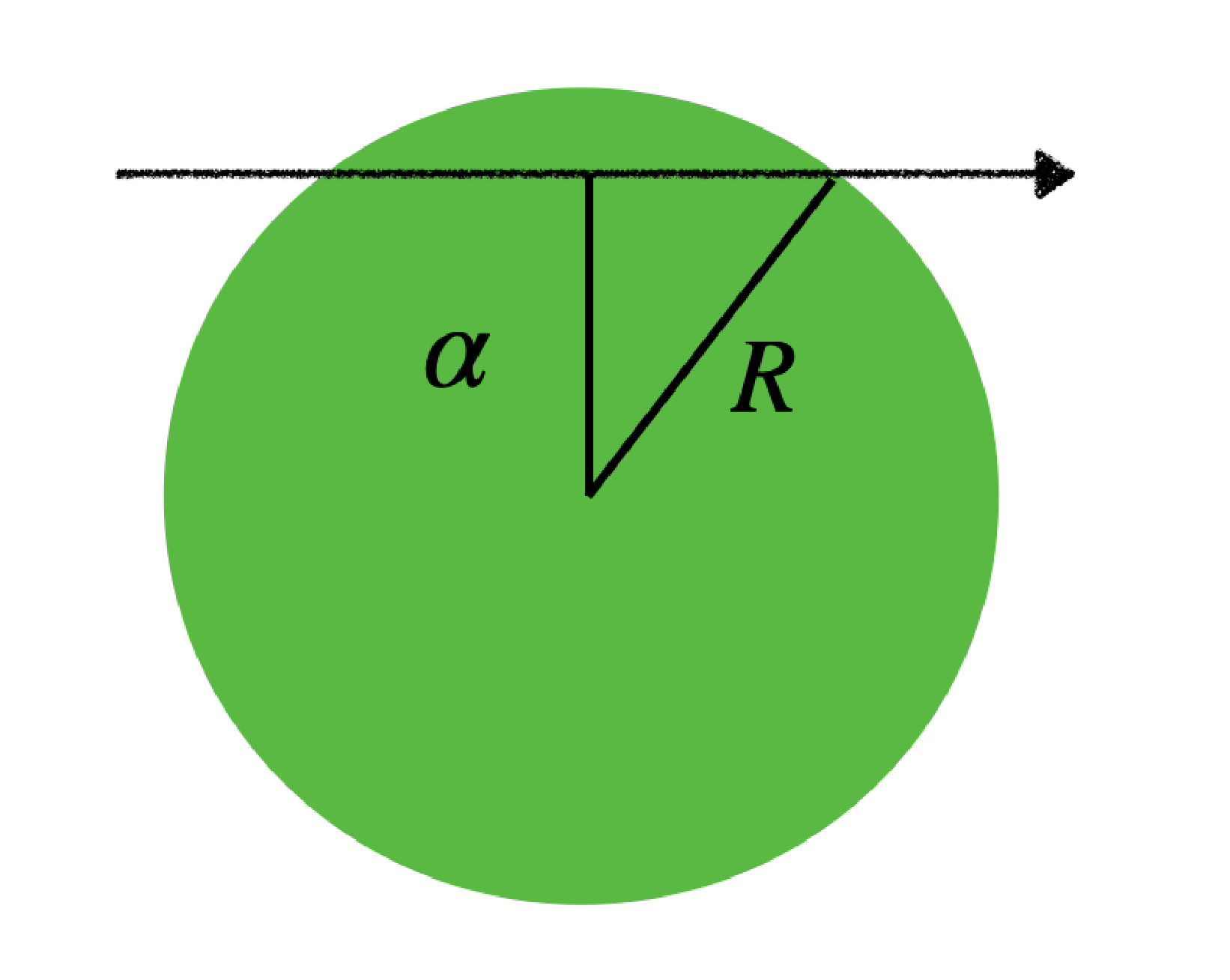}
    \caption{Schematic picture of a minihalo.}
    \label{fig:minihalo}
\end{figure}

\begin{figure}
    \centering
    \includegraphics[width=0.8\hsize]{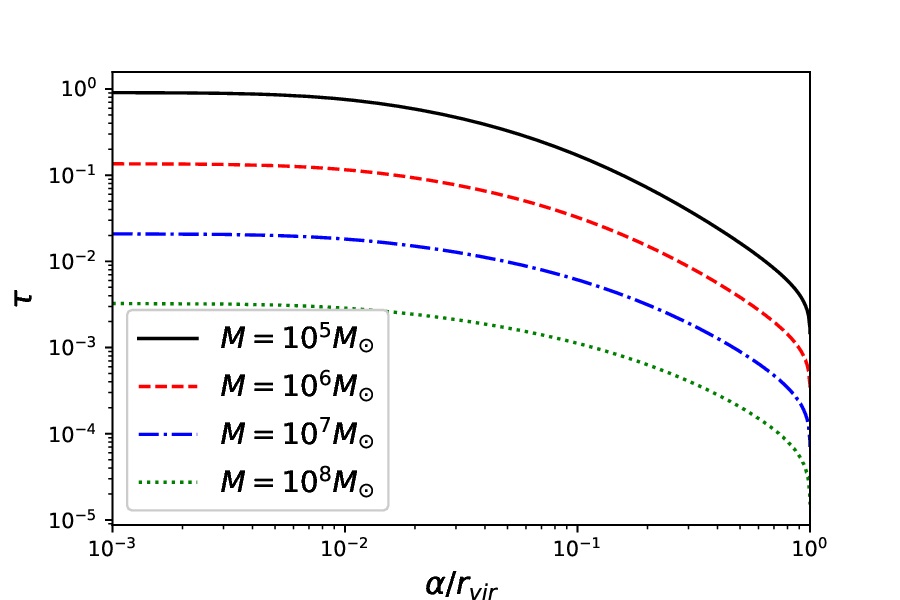}
    \caption{Optical depth as a function of $\alpha/r_{\mathrm{vir}}$ for a given minihalo mass.}
    \label{fig:alpha_tau}
\end{figure}

The optical depth to 21cm absorption lines by neutral hydrogen gas in a minihalo with mass $M$ (at a frequency $\nu$ and at impact parameter $\alpha$) is given by \citep{2002ApJ...579....1F}

\begin{equation}
\begin{aligned}
\tau(\nu, M, \alpha)=& \frac{3 h_{\mathrm{p}} c^3 A_{10}}{32 \pi k_{\mathrm{B}} \nu_{21}^2} \int_{-R_{\max }(\alpha)}^{R_{\max }(\alpha)} d R \frac{n_{\mathrm{HI}}(r)}{T_{\mathrm{S}}(r) \sqrt{\pi} b} \\
& \times \exp \left(-\frac{v^2(\nu)}{b^2}\right),
\end{aligned}
\end{equation}
where $r^2=\alpha^2+R^2$ and $R_{\mathrm{max}}$ is the maximum radius of the halo at $\alpha$. In Fig.\ref{fig:minihalo}, we show the schematic picture of a minihalo. The exponential factor represents Doppler broadening with $v(\nu)=c(\nu-\nu_{21}))/\nu_{21}$ ($\nu_{21}$=1.4GHz), and $b=\sqrt{2 k_B T_{\mathrm{vir}} / m_p}$ is the velocity dispersion. Here, $n_{\mathrm{HI}}$ is the number density of neutral hydrogen atoms in the minihalo. A smaller impact parameter results in a larger
optical depth due to a larger column density despite a larger
spin temperature.  As we mentioned above, the spin temperature approaches the virial temperature in the inner regions of a minihalo; thus, a smaller minihalo has a smaller spin temperature and this results in larger optical depth at a fixed impact parameter \cite{2014PhRvD..90h3003S}. In Fig.\ref{fig:alpha_tau}, we show the optical depth as a function of $\alpha/r_{\mathrm{vir}}$ for a given minihalo mass. From this figure, we can see that the optical depth increases in the inner part of the minihalo. This shows that the neutral hydrogen atom is more concentrated in the inner part of the minihalo. We also see that larger minihalos have a larger optical depth. Roughly speaking, we can estimate $n_{\mathrm{HI}} \sim M / r_{\mathrm{vir}}^2$ and $T_S \sim T_{\mathrm{vir}}$. As $r_{\mathrm{vir}} \propto M^{1 / 3}$ and $T_{\mathrm{vir}} \propto M^{2 / 3}$, we find $n_{\mathrm{HI}} \propto M^{1 / 3}$ and $T_{S} \propto M^{2 / 3}$; thus, this leads to $\tau \propto M^{-1/3}$, pointing to smaller optical depth for larger massive halos.

\subsection{Abundance of the 21cm absorption lines}

\begin{figure}
    \centering
    \includegraphics[width=0.8\hsize]{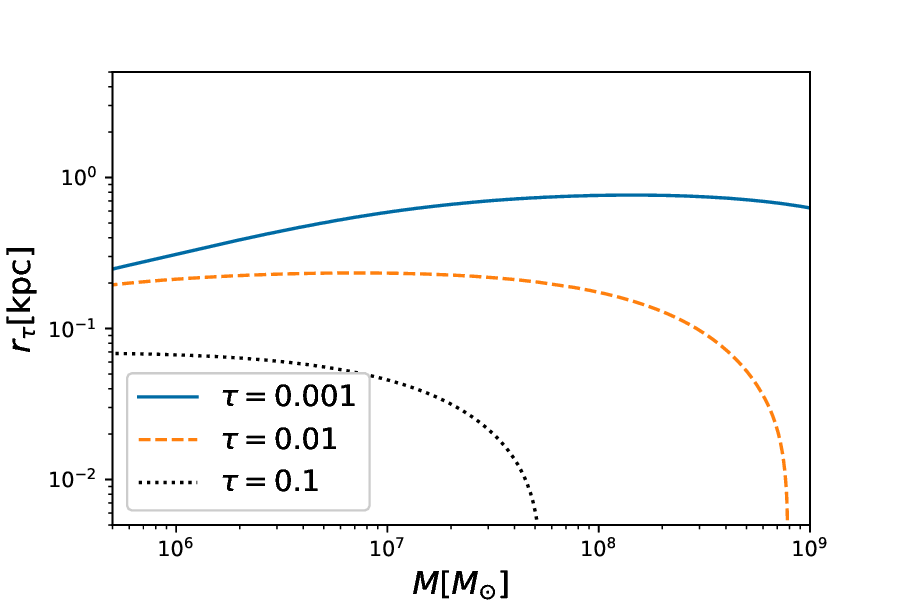}
    \caption{Maximum impact parameter $r_{\mathrm{tau}}$ as function of minihalo mass for given optical depth.}
    \label{fig:mass_rtau}
\end{figure}

In order to evaluate the number of 21cm absorption lines from minihalos in the observed spectrum per redshift, we calculate the abundance of the 21cm line absorption as 

\begin{equation}
\frac{d N(>\tau)}{d z}=\frac{d r}{d z} \int_{M_{\min }}^{M_{\max }} d M \frac{d N}{d M} \pi r_\tau^2(M, \tau),
\label{eq:abundance}
\end{equation}
where $dN/dM$ is the halo mass function which represents the comoving number density of collapsed dark matter halos with a mass between $M$ and $M+dM$; $dr/dz$ is the comoving line element; and $r_\tau(M, \tau)$ is the maximum impact parameter in the comoving unit that gives an optical depth larger than $\tau$. In Fig.\ref{fig:mass_rtau}, we show $r_\tau(M, \tau)$ as a function of the mass of minihalos for a given optical depth. When we focus on a specific minihalo mass, we can see that the maximum impact parameter $r_\tau(M, \tau)$ becomes smaller as the optical depth increases. This is because the optical depth increases in the line of sight direction passing through the inner part of a minihalo as seen by Fig.\ref{fig:minihalo}; thus, the maximum impact parameter becomes smaller for larger optical depth. In our study, we focus on the 21cm forest generated by neutral hydrogen atoms in minihalos. The upper bound of the minihalo mass range of our interest corresponds to the minimal halo mass for which the stars can be formed and the lower bound is the Jeans scale (or a ``filtering scale'', which is essentially the time-averaged Jeans scale, by taking into account the time dependence of the Jeans scale \cite{2000ApJ...542..535G}). Our maximum minihalo mass scale can be bigger in the presence of relative velocity because relative velocity can cause the baryons to overshoot the dark matter halo retarding their collapse. The numerical simulations show that the relative velocity would not change the minimum mass required for the star formation so significantly (typically by less than a factor 2) \cite{Stacy:2010gg,Maio:2010qi,Greif:2011iv,Fialkov:2011iw} . Hence, we conservatively use the typical upper-bound minihalo mass corresponding to the virial temperature $T_{\rm vir} > 10^4$ K which is often adopted in standard discussions to estimate the minihalo contribution to the 21cm forest without the streaming velocity,

\begin{equation}
M_{\max }(z)=3.95 \times 10^{7}\left(\frac{\Omega_{m} h^{2}}{0.15}\right)^{-1 / 2}\left(\frac{1+z}{10}\right)^{-3 / 2} M_{\odot}.
\label{eq:maximum}
\end{equation}

Our lower-bound minihalo mass scale can also increase due to the anisotropic pressure caused by the streaming velocity \cite{2011MNRAS.418..906T,Naoz:2012fr,Fialkov:2014rba}. We set the minimum mass using the baryonic Jeans mass, which takes into account the effect of the relative velocity. The Jeans scale $k_{J}\propto 1/c_s(z)$, which is inversely proportional to the gas sound speed, can be influenced by the stream velocity. We accordingly consider the effective sound speed by taking account of the relative velocity $v_{\mathrm{eff}}=\sqrt{c_s^2(z)+v_{bc}^2(z)}$, so the effective Jeans scale $k_{\mathrm{J,eff}}\propto 1/v_{\mathrm{eff}}$ \cite{Stacy:2010gg, 2011ApJ...735...25N}. The corresponding Jeans mass can be estimated as

\begin{equation}
M_J=\frac{4 \pi \bar{\rho}}{3}\left(\frac{5 \pi k_B T_{\mathrm{IGM}}}{3 G \bar{\rho} m_p \mu}\right)^{3 / 2}\left( \frac{\sqrt{c_\mathrm{s}^2+v_{\mathrm{bc}}^2}}{c_\mathrm{s}}\right)^3 
\label{eq:minimum}
\end{equation}
where $\bar{\rho}$ is the total mass density including dark matter and $T_{\mathrm{IGM}}$ is the temperature of the IGM. We set $T_{\mathrm{IGM}}=2$ K at $z=10$, which satisfies the current lower bound of the IGM temperature obtained by the HERA experiment\cite{2022ApJ...924...51A}. We use Eqs. (\ref{eq:maximum}) and (\ref{eq:minimum}) as the upper and lower minihalo mass scales $M_{\mathrm{max}}$ and $M_{\mathrm{min}}$, respectively, in estimating the 21cm signals.

\section{Results}
\label{Sec4}
\begin{figure}
    \centering
    \includegraphics[width=0.8\hsize]{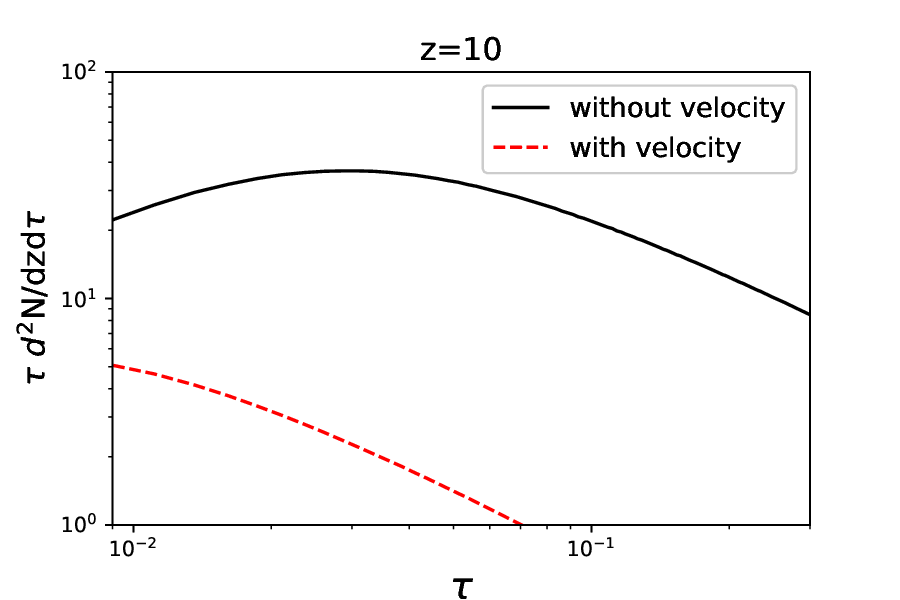}
    \caption{Abundance of the 21cm absorption lines at $z$=10 without relative velocity(solid line) and with relative velocity(dashed line).}
    \label{fig:21cm_forest_z10}
\end{figure}

In Fig.\ref{fig:21cm_forest_z10}, we show the abundance of the 21cm absorption lines at $z=10$ as a function of the optical depth per redshift interval along a line of sight and per optical depth. We show the abundance of the 21cm absorption lines with and without the relative velocity for illustration. The abundance of the 21cm absorption lines without the relative velocity is smaller than in the case that does not include the relative velocity by around 1 order of magnitude. In the absence of the relative velocity, the abundance of the 21cm absorption lines per line-of-sight direction is around $O(10)$ at $\tau \lesssim 0.1$, whereas, in the presence of relative velocity, it decreases to around $O(1)$. The impacts of relative velocity on the 21cm forest appear via the suppression of the mass function (or matter power spectrum) and the enhancement of the Jeans mass. As shown by Fig.\ref{fig:MF}, the presence of relative velocity suppresses the halo mass function and, as shown in Eq.(\ref{eq:abundance}), the suppression of the mass function reduces the number of 21cm absorption lines. 
As shown in Eq.(\ref{eq:minimum}), the Jeans mass is roughly 20 times larger due to the relative velocity. 
The Jeans mass affects the minimum minihalo mass which contributes to the abundance of 21cm absorption lines. Such a change can substantially reduce the 21cm signals because the abundance of the 21cm absorption lines is more significantly affected by the smaller minihalos with a smaller spin temperature and a larger optical depth, rather than by the bigger minihalos \cite{2014PhRvD..90h3003S}.

\begin{figure}
    \centering
    \includegraphics[width=0.8\hsize]{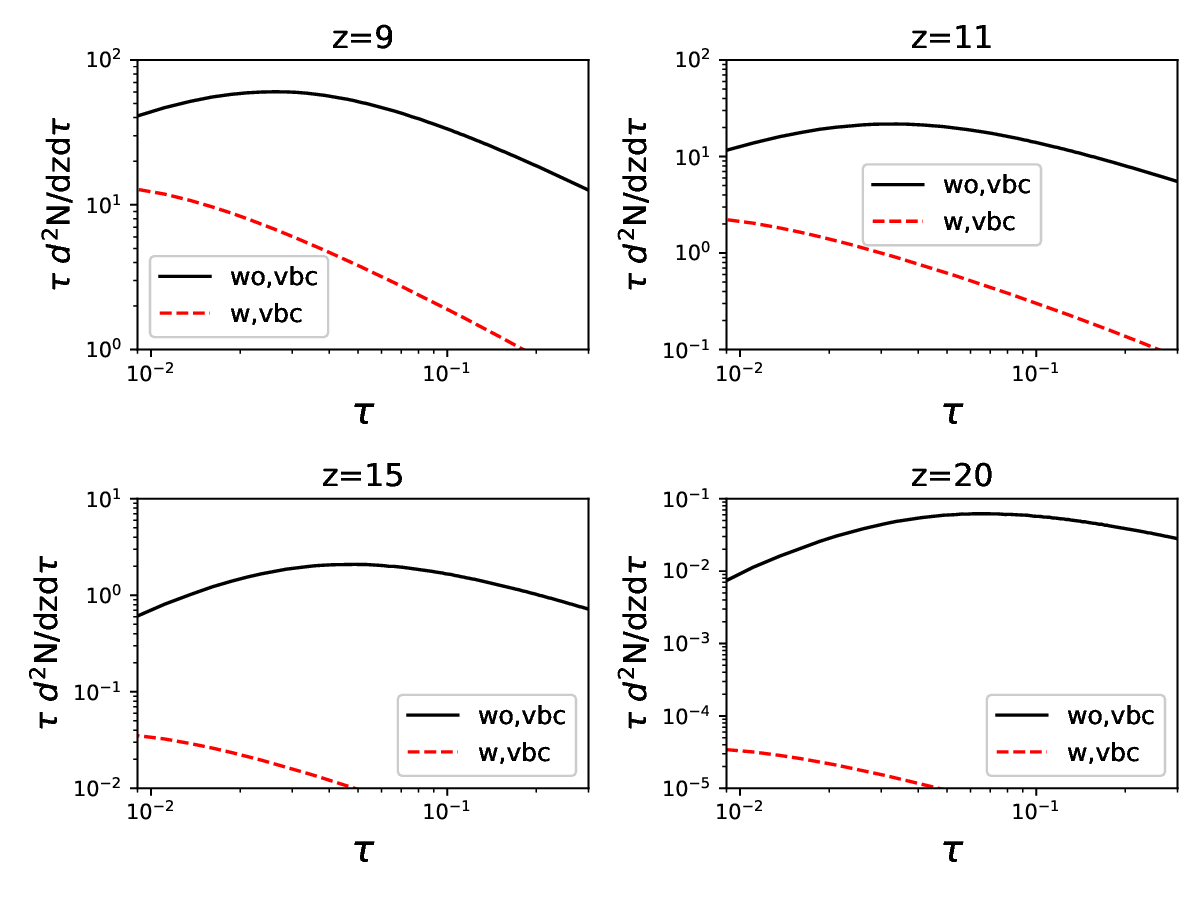}
    \caption{Abundance of the 21cm absorption lines at different redshifts.}
    \label{fig:21cm_forest_zevolution}
\end{figure}

In Fig.\ref{fig:21cm_forest_zevolution}, we show the abundance of 21cm absorption lines at different redshifts. 

At a higher redshift, the relative suppression of the 21cm absorption line abundance is more prominent. One reason for this is the halo mass function. At a higher redshift, the relative velocity $\propto (1+z)$ is bigger and a smaller halo has a less-deep gravitational potential well. Another reason for the abundance is the Jeans mass. In the absence of X-ray heating, the IGM gas temperature scales as $\propto (1+z)^2$.[Eq.(\ref{eq:minimum})]. 
 
At a lower redshift, the number of halos increases, and the gas temperature (Jeans mass) decreases. Thus, the number of 21cm absorption lines increases. At $z=11$ and $z=15$, the abundance of the 21cm absorption lines is $O(1)-O(10)$ in the absence of the relative velocity. Taking the relative velocity into account, we can see from the figures that the abundance of the 21cm absorption lines decreases while keeping $O(1)$ at a lower $\tau$ (up to $\sim 0.03)$ at $z=11$. At $z=20$, even in the presence of the relative velocity, the number of 21cm absorption lines is less than $O(0.1)$ for any optical depth and it becomes less than $O(10^{-4})$ if the effect of relative velocity is included.

\begin{figure}
    \centering
    \includegraphics[width=0.8\hsize]{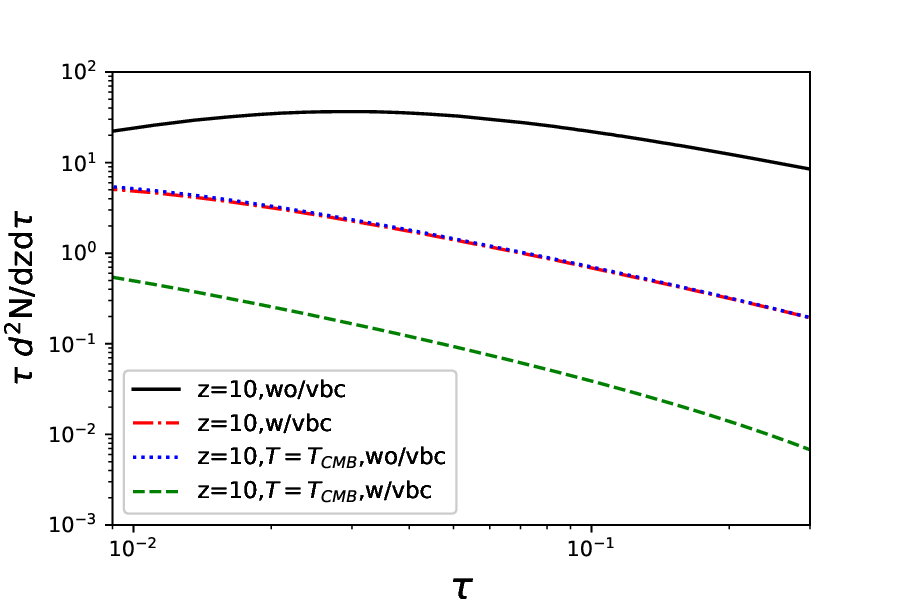}
    \caption{Abundance of the 21cm absorption lines for different IGM temperatures. We show the number of 21cm absorption lines for no relative velocity($T_{\mathrm{IGM}}=2\mathrm{K}$) (solid line), including relative velocity($T_{\mathrm{IGM}}=2\mathrm{K}$ )(dashed-dot line), no relative velocity ($T_{\mathrm{IGM}}=T_{\mathrm{CMB}}$)(dotted line) and including relative velocity($T_{\mathrm{IGM}}=T_{\mathrm{CMB}}$ )(dashed line).}
    \label{fig:21cm_forest_temperature}
\end{figure}

For illustration, we have shown the 21cm absorption abundance where the minimum minihalo mass is determined by the Jeans mass with $T_{\mathrm{IGM}}=2\mathrm{K}$ at $z=10$ and the kinetic temperature evolves as $\propto (1+z)^2$. We next show the temperature dependence of the number of 21cm absorption lines. Fig.\ref{fig:21cm_forest_temperature} shows the abundance of the 21cm absorption when the IGM temperature is equal to the CMB temperature at $z=10$ ($T_{\mathrm{IGM}}=30\mathrm{K}$). When the kinetic temperature becomes larger, the Jeans mass becomes larger. In our formalism, the minimum minihalo mass is determined by the Jeans mass. The contributions to 21cm absorption lines mainly come from smaller minihalos (which possess a larger optical depth than the larger minihalos \cite{2014PhRvD..90h3003S}), and a larger kinetic temperature of the IGM reduces the abundance of 21cm absorption lines. In Fig.\ref{fig:21cm_forest_temperature}, for $T_{\mathrm{IGM}}=T_{\mathrm{CMB}}$, we can see that the abundance of the 21cm forest reduces to less than $\mathcal{O}(10)$ in both cases with and without the relative velocity. In particular, the abundance of the 21cm absorption lines is less than 1 if we take the relative velocity into account. We also find that the abundance of the 21cm absorption lines including the relative velocity with $T_{\mathrm{IGM}}=2\mathrm{K}$ is very similar to that of the 21cm absorption lines not including the relative velocity with $T_{\mathrm{IGM}}=T_{\mathrm{CMB}}$. Although this behavior is just a coincidence in this example, it illustrates that raising the IGM temperature and including the relative velocity generates the degeneracy for the abundance of the 21cm absorption lines.

\begin{figure}
    \centering
    \includegraphics[width=0.8\hsize]{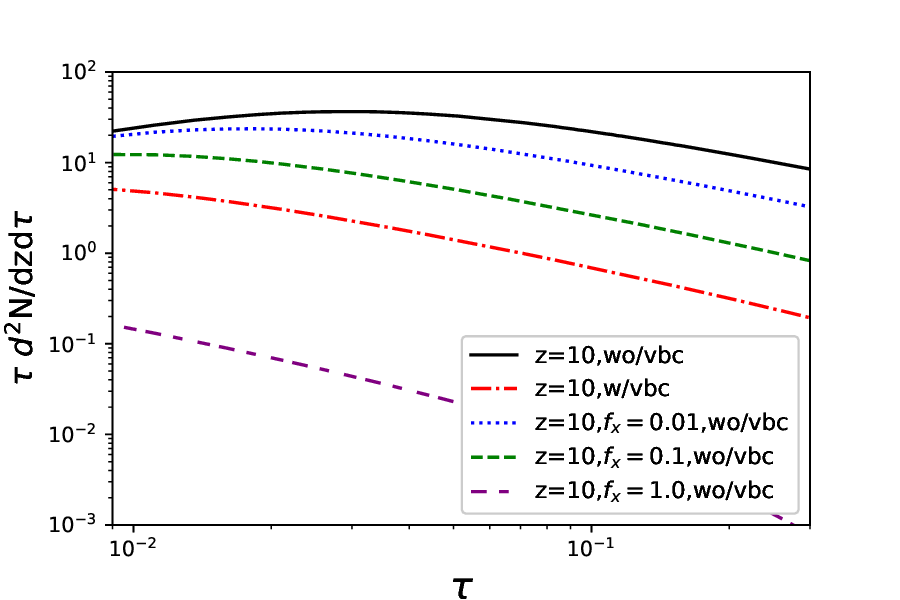}
    \caption{Abundance of the 21cm absorption lines for different X-ray heating efficiency $f_X$. To evaluate the pure impact of $f_X$ on the 21cm absorption lines, we do not include the effect of relative velocity for different $f_X$ cases. For comparison, we also show the abundance of the 21cm absorption lines with and without relative velocity for $T_{\mathrm{IGM}}=2\mathrm{K}$.}
    \label{fig:21cm_forest_temperature2}
\end{figure}

In addition, we also consider the effect of X-ray heating on the 21cm forest in Fig.\ref{fig:21cm_forest_temperature2}. The energy injections into the IGM from X-ray sources are associated with X-ray efficiency, which is parameterized as $f_X$ (See e.g., Refs.\citep[e.g.][]{2006MNRAS.370.1867F,2013MNRAS.431..621M,2021MNRAS.506.5818S}). A higher $f_X$ means that the X-ray heating is more efficient and the IGM is heated much more. As the IGM is heated, the Jeans mass of the minihalos is increased. This means that a higher $f_X$ reduces the number of minihalos in which the neutral hydrogen atom contributes to the production of the 21cm forest and thus reduces the number of 21cm absorption lines. In particular, the abundance of the 21cm absorption lines at larger optical depth is reduced because the larger Jeans mass reduced the smaller mass of minihalos which mainly contribute to producing the 21cm absorption lines at larger optical depth.  From Fig.\ref{fig:21cm_forest_temperature2}, we can see the impact of $f_X$ on the 21cm absorption lines. To evaluate the effect of $f_X$, we do not include the effect of relative velocity for different $f_X$ cases. As $f_X$ increases, the abundance of the 21 cm absorption lines decreases, in particular, for $f_X$=1, the abundance of the 21 cm absorption lines is less than 1 at any optical depth. For comparison, we also show the abundance of the 21cm absorption lines with and without relative velocity for $T_{\mathrm{IGM}}=2\mathrm{K}$. Comparing the case with the relative velocity at $T_{\mathrm{IGM}}=2\mathrm{K}$ and with $f_X$=0.01 and 0.1, we can see that the effect of relative velocity reduces the number of 21cm line absorption lines more than the case with  $f_X$=0.01 and 0.1.

\section{Summary and Discussion}
\label{Sec5}
In this paper, we investigated the possible impact of dark matter-baryon relative velocity on the 21cm forest. The relative velocity generates additional anisotropic pressure, which suppresses the structure formation and consequently reduces the number of 21cm absorption lines produced by neutral hydrogen atoms in the minihalos. In this work, we focus on the 21cm forest produced by neutral hydrogen atoms inside the minihalos. The neutral hydrogen atoms in the diffuse IGM and the structure of the filament also contribute to the optical depth of the 21cm forest; however, we neglect them in this study because the impact of relative velocity on the 21cm forest appears through the formation of the minihalos, and also because the contributions from diffuse IGM and the filaments to the 21cm forest are small due to their small number density of neutral hydrogen atoms compared to those of the minihalos (See, e.g., Refs. \citep[][]{2002ApJ...577...22C,2002ApJ...579....1F}). Compared with the 21cm emission lines from diffuse IGM, the advantage of the 21cm forest observation is to avoid the diffuse foregrounds such as the synchrotron emission from our Galaxy because the 21cm forest observation utilizes the 21cm absorption lines in the spectra from bright radio sources. On the other hand, radio background sources have a disadvantage because the 21cm forest strongly relies on the radio bright sources at a high redshift. According to \cite{2021MNRAS.506.5818S}, we can estimate the required minimum brightness of a radio background source for the detection of the 21cm forest by

\begin{equation}
\begin{aligned}
S_{\min }=& 17.2 \mathrm{mJy}\left(\frac{0.01}{1-F_{21, \mathrm{th}}}\right)\left(\frac{\mathrm{S} / \mathrm{N}}{5}\right)\left(\frac{5 \mathrm{kHz}}{\Delta \nu}\right)^{1 / 2}\\
&\times \left(\frac{1000 \mathrm{hr}}{t_{\mathrm{int}}}\right)^{1 / 2}\left(\frac{600 \mathrm{~m}^2 \mathrm{~K}^{-1}}{A_{\mathrm{eff}} / T_{\mathrm{sys}}}\right)
\label{eq:Smin}
\end{aligned}
\end{equation}
where $F_{21,\mathrm{th}}=e^{-\tau}$ is 21cm transmission for 21cm optical depth $\tau$, $\Delta \nu$ is a frequency resolution, $A_{{\rm eff}}/T_{{\rm sys}}$ is the ratio of an effective collecting area and a system temperature and $t_{{\rm int}}$ is the observation time. In Eq.(\ref{eq:Smin}), we normalize each quantity by the SKA-like specifications.

Recently, over 280 quasars have been reported at $z>5.7$ (see, e.g., Ref.\citep[][]{sarah_e_i_bosman_2021_5510200}). Based on the discovery of luminous quasars at high redshift by near-infrared observations, some previous works estimate the number of luminous quasars at high redshift by extrapolating the observed number density of quasars (see, e.g., Refs.\citep[][]{2016ApJ...833..222J,2019ApJ...884...30W,2019BAAS...51c.121F}). According to their estimation, we predict there will be only $\sim 1$ quasars at $z\sim 9$. We have also discovered quasars through radio observations (see, e.g., Ref.\citep[][]{refId0}). Some of these quasars are radio-loud active galactic nuclei (AGNs) (see, e.g., Refs.\citep[][]{ban2015,ban2018,2020A&A...635L...7B}) such as the blazar PSO J030947.49+271757.31 with a flux density $S_{147\mathrm{MHz}}=64.2\pm 6.2 \mathrm{mJy}$ \citep{2020A&A...635L...7B}. Some previous works have also estimated the number of radio-loud quasars based on extrapolations of the observed radio luminosity function to high redshift, and they predicted the number of quasars is rough of order $O(1)-O(10)$ at $z=10$ in the case where the redshift evolution of the number density of quasars is steep\citep{Haiman_2004,Xu_2009,2015aska.confE...6C}. However, the estimation of the number density of the quasars depends on the redshift evolution of the luminosity function, survey area, and observation time. Thus, the estimation of the number of radio-loud quasars at high redshift has large uncertainties and is still under debate.


Several experiments, such as the Low-Frequency Array (LOFAR) Two-metre Sky Survey (LoTSS) 
 (see, e.g., Refs.\citep[][]{2017A&A...598A.104S,2019A&A...622A...1S,2022A&A...659A...1S}), the Giant Metrewave Radio Telescope (GMRT) all-sky radio survey at 150 MHz \citep{2017A&A...598A..78I}, and the Galactic and Extragalactic All-sky Murchison Widefield Array survey (GLEAM) \citep{2015PASA...32...25W}, are projected to detect hundreds of bright radio sources at $z>$6. These observational surveys and theoretical predictions support future 21cm forest studies. 

As seen in Fig.\ref{fig:21cm_forest_temperature}, the 21cm forest generated by neutral hydrogen atoms in the minihalos strongly depends on the thermal state of the IGM because the IGM temperature regulates the minimum mass of minihalos sensitive to the estimated number of 21cm absorption lines. Although the HERA experiment currently puts the lower limits on the IGM temperature at $z\sim 8$\citep{2022ApJ...924...51A}, the uncertainty of the thermal state of the IGM at a high redshift is still large. Nevertheless, it is worth mentioning that the higher IGM temperature can further suppress the number of 21cm absorption lines.

So far, we have assumed the conventional $\Lambda$CDM cosmology. The 21cm observations can be useful to probe different cosmological models. For instance, if we consider the warm dark matter or the axionlike ultralight particles, the number of 21cm absorption lines can be reduced due to the free streaming or``"quantum pressure" \citep{2014PhRvD..90h3003S,2020PhRvD.101d3516S}. On the other hand, if the axionlike particles are generated due to spontaneous symmetry breaking after inflation (the so-called the post inflation Peccei-Quinn (PQ) symmetry breaking scenario), the isocurvature perturbations can be generated, which can dominate at small scales and enhance the 21cm forest signals \citep{2020PhRvD.102b3522S,2021JCAP...04..019K,2022JCAP...08..066K}. The application of the 21cm forest to the study of dark matter, such as axionlike cold dark matter and warm dark matter in the presence of dark matter-baryon relative velocity is also left for future work.

While our paper focuses on the suppression of small-scale structures due to the relative velocity, we also mention that the relative velocity can potentially help in the formation of massive stars.
For instance, recent cosmological simulations have shown that relative velocity can play an important role in forming massive stars and those massive stars become the seeds for supermassive black holes (SMBH) \cite{2017Sci...357.1375H}. Supersonic relative velocity prevents gas cloud formation until rapid gas condensation is triggered in a protogalactic halo. When the gas cloud becomes gravitationally unstable, it directly collapses, and the supermassive star, which is $M \sim $34,000 $M_{\odot}$, can form. Such supermassive stars can become the seed for SMBHs. The phenomenology relevant to the relative velocity is rich and further study is warranted.  

The 21cm forest can be a useful probe for small-scale structures, which can go even go beyond $k\gtrsim 10 \mathrm{Mpc}^{-1}$, and we illustrate its usefulness by demonstrating that it is indeed sensitive to the dark matter-baryon relative velocity. Further application of 21cm forest observations to the small-scale structures, such as its usefulness to study the subhalo structures, would also be worth exploring, and this is left for future work \cite{Kadota:2022uvg}.

\begin{acknowledgments}
 This work is supported by the National SKA Program of China (Grant No.~2020SKA0110401), NSFC (Grant No.~12103044), JSPS Grant No. 18K03616 and 17H01110,21H04467, JST AIP Acceleration Research Grant No.JP20317829, and JST FOREST Program No. JPMJFR20352935. K.K. thanks the Yukawa Institute for Theoretical Physics and the Nagoya University Cosmology Group for their hospitality during the completion of this work.
\end{acknowledgments}


\bibliography{ref}


\end{document}